\documentclass[12pt]{iopart}

\usepackage{iopams}
\usepackage{graphicx}

\begin{document}

\title[ ]{Perspectives for a mixed two-qubit system with
binomial quantum states }

\author[Mahmoud Abdel-Aty]{Mahmoud Abdel-Aty\footnote[3]{E-mail: abdelatyquant@yahoo.co.uk}
}
\address{Mathematics Department, Faculty of Science, South Valley University, 82524 Sohag, Egypt.
}

\begin{abstract}

The problem of the relationship between entanglement and two-qubit systems
in which it is embedded is central to the quantum information theory. This
paper suggests that the concurrence hierarchy as an entanglement measure
provides an alternative view of how to think about this problem. We consider
mixed states of two qubits and obtain an exact solution of the
time-dependent master equation that describes the evolution of two two-level
qubits (or atoms) within a perfect cavity for the case of multiphoton
transition. We consider the situation for which the field may start from a
binomial state. Employing this solution, the significant features of the
entanglement when a second qubit is weakly coupled to the field and becomes
entangled with the first qubit, is investigated.
We also describe the response of the atomic system as it varies between the
Rabi oscillations and the collapse-revival mode and investigate the atomic inversion and the Q-function. We
identify and numerically demonstrate the region of parameters where
significantly large entanglement can be obtained. Most interestingly, it is
shown that features of the entanglement is influenced significantly when the
multi-photon process is involved. Finally, we obtain illustrative examples
of some novel aspects of this system and show how the off-resonant case can
sensitize entanglement to the role of initial state setting.
\end{abstract}

\submitto{\JOB}

\maketitle

\section{Overview}

As a lot of researchers suggest [1-5], entanglement is one of the
most mysterious aspects of quantum physics. Although it is by now
possible to verify the predictions of entanglement theory in a
variety of experiments [2], there remains a considerable gap
between the formal definition of entanglement and the observable
effects that are associated with this property. A major thrust of
current research is to find a quantitative measure of entanglement
for general states. For the experiments in the newest fields of
physics, quantum computing, quantum communication, and quantum
cryptography [6-13] the quantitative analysis of the multi-qubit
or ion is of substantial interest [14]. The dipole-dipole
interaction between two atoms can be understood through the
exchange of virtual photons and depends on the transition dipole
moment of the levels involved. It can be characterized by complex
coupling constants, or by their real and imaginary parts, where
the former affect decay constants and the latter lead to level
shifts [15,16]. There is an inherent interest in analytical and
non-perturbative solutions of multi-atom interacting with the
cavity field problems, all the more considering quantum systems
with more than one qubit. One examples of such kind is the system
of two two-level qubits in an electromagnetic field [14-26].
Entanglement of identical qubits is a property dependent on which
single-qubit basis is chosen, as any operation should act on each
identical qubit in the same way. Indeed, individual qubits are
excitations of a quantum field, and the single-qubit basis defines
which set of qubits are used in representing the many-qubit state
 [27]. The analysis of entanglement sharing of the
two-atom Tavis-Cummings model has been discussed in Ref. [28].
Recently, interest has mounted in exploring the quantum system
composed of two qubits interacting with a thermal field [29]. We
have addressed a general two-qubit system in a recent paper [23]
in which an analytical expression for the temporal evolution of
the Pancharatnam phase when the field starts from vacuum state is
given.

This has motivated us to criticize the conception of entanglement
of a two-qubit system in the context of the mixed quantum states.
The main contribution of this paper is to synthesize conceptual
insights that already exist to push forward a more coherent view
of how concurrence as an entanglement measure of a two-qubit
system might make progress in understanding the two-qubit
entanglement. To be more precise, we assume that two two-level
atoms (two qubits) share a bipartite system, taking into account
the multi-photon transition. Another principal aim is to elucidate
the extent to which mixed entangled states can affect the
entanglement. The emphasis being put on the investigation of the
entanglement in a more general situation in which the two atoms
(qubits) share a mixed state, rather than a pure state and we
propose to use the concurrence hierarchy as a measurement of
entanglement [30-32]. The issue of attributing objective
properties to the constituents of a quantum system composed of two
qubits, does not turn out to be a straightforward generalization
of the just analyzed case involving distinguishable qubits, and
the problem of entanglement has to be reconsidered. Entangled
mixed states may arise when one or both qubits of an initially
pure entangled state interact, intentionally or inadvertently,
with other quantum degrees of freedom resulting in a non-unitary
evolution of the pure state into a mixed state. In general it is
known that there are also cases when entangled states are mixed
with other entangled states and where the sum is separable.

The rest of the paper is organized as follows. We devote section 2
to give a brief overview of the binomial states and initial states
setting. In section 3, we present notations and definitions of the
model and its analytical solution to be used in the rest of the
paper. Section 4 is devoted to consider the atomic inversion as
well as the quasiprobability distribution. The concurrence as an
entanglement measure is presented in section 5, followed by a
numerical computation in which we shall examine the influence of
different involved parameters in subsection 5.1. The paper is
closes with the conclusions outlined in section 6.

\section{Binomial state}

The implementation independence in quantum information theory is guaranteed
by the use of Hilbert spaces, states and operations between and on them. It
is not said, what they physically describe in more concrete terms, whether
we are dealing with spins, polarizations, energy levels, qubit numbers, or
whatever you can imagine. The features of nonclassical states visualize
specific aspects of nonclassicality and do not yield a complete
characterization of nonclassicality as a phenomenon on its own. In this
section we give a brief overview of the binomial states [33]. For measuring
the quantum state of the radiation field, balanced homodyning has become a
well established method, it directly measures phase sensitive quadrature
distributions. Alternatively, unbalanced homodyning yields access to
phase-space functions. For some systems, such as a cavity-field mode and a
mode of the quantized motion of a trapped ion, methods have been proposed
that allow for a direct detection of the characteristic functions of the
quadratures. The binomial states are finite linear combinations of number
states {[33]}
\begin{equation}
|\eta, m\rangle =\sum\limits_{n=0}^{\infty }\left[ \left(
\begin{array}{c}
m \\
n
\end{array}
\right) \eta^{n}(1-\eta )^{m-n}\right]^{1/2}|n\rangle ,
\end{equation}
which interpolate between some fundamental sates such as number states and
coherent states, where $m$ is a non-negative integer, $\eta $ is areal
probability $(0<\eta <1)$ and $|n\rangle $ is a number state of the
radiation field. Such states have been studied in great detail in the
literature (see e.g. Ref. {[34]}). The binomial states have the properties
\begin{equation}
|\eta ,m\rangle =\left\{
\begin{array}{c}
|m\rangle \\
|0\rangle \\
|\alpha \rangle
\end{array}
\right.
\begin{array}{c}
\eta \rightarrow 1 \\
\eta \rightarrow 0 \\
\eta \rightarrow 0,
\end{array}
\begin{array}{c}
\\
\\
m\rightarrow \infty ,\quad \eta m=\alpha^{2}.
\end{array}
\end{equation}
In the theory of open system or the reduction theory, one often
considers two subsystem $\mathrm{A}$ and $\mathrm{F}$ represented
by Hilbert space. Let $\frak{S}_{j},\,\,(j=A,F)$ be state spaces
(the set of all density operators) and $\frak{S}_{A}\otimes
\frak{S}_{F}$ denotes the state space in the composite system.
Here, we assume that, before entering the cavity, the field
initially in a binomial state such as
\begin{equation}
\varpi =|\eta ,m\rangle \langle \eta ,m|\in \frak{S}_{F}.
\end{equation}
In pure-state quantum mechanics the state of the system is usually
represented by a normalized wavefunction, which is a unit vector
in a Hilbert space. Entangled mixed states may arise when the
qubits of an initially pure entangled state interact,
intentionally or inadvertently, with other quantum degrees of
freedom resulting in a non-unitary evolution of the pure state
into a mixed state. In general it is known that there are also
cases when entangled states are mixed with other entangled states
and where the sum is separable. The usual interpretation of mixed
states, is that their creation involves irreversibly destroying
information [35,36]. This has interesting consequences concerning
entanglement theory, since there, the irreversibility is often
associated with the fact that one is dealing with mixed states. We
assume that, the two two-level qubits initially prepared in the
mixed states enter the cavity whose single mode under
consideration, prior to the interaction with the qubits, is in the
binomial state $\varpi .$ Thus, the atomic density matrix is of
the diagonal form
\begin{equation}
\rho_{i}^{a_{i}}(0)=\cos^{2}\theta_{i}\left| e_{i}\right\rangle
\left\langle e_{i}\right| +\sin^{2}\theta_{i}|g_{i}\rangle \langle
g_{i}|\in \frak{S}_{A_{i}},
\end{equation}
where $\rho_{i}^{a_{i}}(0)$ is the atomic density matrix for the $i^{
\underline{th}}$ qubit. The initial state of the two qubits system can be
written as
\begin{equation}
\rho^{a}(0)=\rho_{1}^{a}(0)\otimes \rho_{2}^{a}(0)\in \frak{S}_{A}.
\end{equation}
\smallskip It is well known that mixed states can be realized by an ensemble
of pure states in an infinite number of ways. The determination of
the separability of a state and the determination of its
entanglement have in common that a particular realization of a
state has to be found such that some property holds for all pure
states in that realization. In order to find this optimal
realization, it is of considerable interest to have a
mathematically elegant way of generating all possible realizations
of a state [31].

\section{Model for pair of qubits}

We consider a mechanism through which a system of two qubits can
be entangled in a cavity field. We assume that the qubits are
modeled by two-level systems having multiphoton transition, which
is a micromaser system [21]. The micromaser is an experimental
realization of the idealized system of a two-level qubit
interacting with a second quantized single-mode of the
electromagnetic field [22]. In this section we consider a
theoretical model which differs from the standard micromaser
set-up in that instead of a single qubit we have assumed a pair of
qubits interacting with a single mode of the cavity field (see
figure 1). In the dipole and rotating wave approximation, one can
write $(\hbar =1)$
\begin{figure}[tbph]
\begin{center}
\includegraphics[width=14cm]{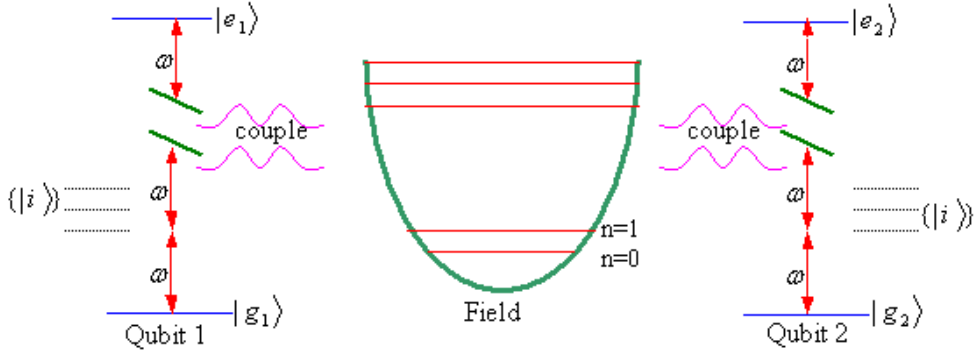}
\end{center}
\caption{Schematic of the multiphoton transition in the interaction of a
pair of two two-level qubits with a single-mode radiation field. In the
shown process we denote by $|e_{j}\rangle $($|g_{j}\rangle $) the $j^{
\underline{th}}$ qubit's upper (lower) level and $\{|i\rangle \}$
are the virtual levels. }
\end{figure}
\begin{eqnarray}
\hat{H} &=&\omega \hat{a}^{\dagger }\hat{a}+\omega_{1}\left[
S_{ee}^{(1)}-S_{gg}^{(1)}\right] +\omega_{2}\left[ S_{ee}^{(2)}-S_{gg}^{(2)}
\right] +\hat{a}^{\dagger }\hat{a}\{\beta_{1}^{(1)}\hat{\sigma}_{-}^{(1)}
\hat{\sigma}_{+}^{(1)}
\nonumber
\\
&&+\beta_{1}^{(2)}\hat{\sigma}_{-}^{(2)}\hat{\sigma}_{+}^{(2)}+\beta
_{2}^{(1)}\hat{\sigma}_{+}^{(1)}\hat{\sigma}_{-}^{(1)}+\beta_{2}^{(2)}\hat{
\sigma}_{+}^{(2)}\hat{\sigma}_{-}^{(2)}\}\Theta (k)+\gamma_{1}S_{eg}^{(1)}
\hat{a}^{k}
\nonumber
\\
&&+\gamma_{1}^{\ast }S_{ge}^{(1)}\hat{a}^{\dagger k}+\gamma_{2}S_{eg}^{(2)}
\hat{a}^{k}+\gamma_{2}^{\ast }\hat{a}^{\dagger k}S_{ge}^{(2)}.
\end{eqnarray}
We denote by $S_{lm}^{(i)}$ the atomic operators for the $i^{{\underline{th}}
}$ qubit. $\hat{a}$ and $\hat{a}^{\dagger }$ are field operators
corresponding to annihilation and creation of photons in the cavity mode. We
denote by $\gamma_{i}$ the coupling constant for the $i^{{\underline{th}}}$
qubit, $\omega $ is the field frequency and $\omega_{i}$ is the atomic
frequency for the $i^{\underline{th}\ }$qubit. The parameter $\Theta (k)$ is
defined such that
\begin{equation}
\Theta (k)=\left\{
\begin{array}{c}
0\qquad k=1 \\
\qquad 1\qquad k>1\qquad
\end{array}
\right.
\end{equation}
If $k$ is larger than unity, in a consistent physical treatment one should
take into account field-induced level shifts which are proportional to the
number of photons in the field mode. Here we denote by $\beta_{1}^{(i)}$ and $
\beta_{2}^{(i)}$ the intensity-dependent Stark shifts to the two levels of the $i^{\underline{th}}$ qubit, that are due to the virtual transitions to the intermediate level.

We write the Hamiltonian $\hat{H}$ into two mutually commuting parts $\hat{H}
=\hat{H}_{0}+\widehat{H}_{in},$ where [$\hat{H}_{0},\widehat{H}_{in}]=0,$
\begin{eqnarray}
\hat{H}_{0} &=&\omega \left( \hat{a}^{\dagger }\hat{a}
+S_{ee}^{(1)}+S_{ee}^{(2)}-S_{gg}^{(1)}-S_{gg}^{(2)}\right) ,
\nonumber
\\
\widehat{H}_{in} &=&\hat{a}^{\dagger }\hat{a}\left[ \beta_{1}^{(1)}\hat{
\sigma}_{-}^{(1)}\hat{\sigma}_{+}^{(1)}+\beta_{1}^{(2)}\hat{\sigma}
_{-}^{(2)}\hat{\sigma}_{+}^{(2)}+\beta_{2}^{(1)}\hat{\sigma}_{+}^{(1)}\hat{
\sigma}_{-}^{(1)}+\beta_{2}^{(2)}\hat{\sigma}_{+}^{(2)}\hat{\sigma}
_{-}^{(2)}\right] \Theta (k)
\nonumber
\\
&&+\Delta \left[ S_{ee}^{(1)}+S_{ee}^{(2)}-S_{gg}^{(1)}-S_{gg}^{(2)}\right]
+\gamma_{1}S_{eg}^{(1)}\hat{a}^{k}+\gamma_{1}^{\ast }S_{ge}^{(1)}\hat{a}
^{\dagger k}
\nonumber
\\
&&+\gamma_{2}S_{eg}^{(2)}\hat{a}^{k}+\gamma_{2}^{\ast }\hat{a}^{\dagger
k}S_{ge}^{(2)}.
\end{eqnarray}
The detuning parameter is given by $\Delta =\omega_{i}-k\omega .$ The
continuous map $\mathcal{E}_{t}^{\ast }$ describing the time evolution
between the qubits and the field is defined by the unitary evolution
operator generated by $\hat{H}_{in}$ such that
\begin{eqnarray}
\mathcal{E}_{t}^{\ast } &:&\frak{S}_{A}\longrightarrow \frak{S}_{A}\otimes
\frak{S}_{F},  \nonumber \\
\mathcal{E}_{t}^{\ast }\rho &=&\widehat{U}_{t}\left( \rho^{a}(0)\otimes
\varpi \right) \widehat{U}_{t}^{\ast }.  \label{rho}
\end{eqnarray}
The interaction Hamiltonian, in this case, leads to an exactly solvable time
evolution operator. Resuming our analysis, the time evolution operator can
be written as
\begin{equation}
\widehat{U}_{t}\equiv \exp \left( -\frac{i}{\hbar }\int\limits_{0}^{t}\hat{H}
(t^{\prime })dt^{\prime }\right) .  \label{uni}
\end{equation}
A solution of equation (\ref{rho}) can be written as
\begin{equation}
\mathcal{E}_{t}^{\ast }\rho
=\sum\limits_{i=1}^{4}\sum\limits_{z=1}^{4}\sum\limits_{n=0}^{m}\sum
\limits_{l=0}^{m}\Omega_{iz}(n,v,t)\left| \Psi_{i}\right\rangle
\left\langle \Psi_{z}\right| ,  \label{f-rho}
\end{equation}
where
\begin{eqnarray}
\Omega_{iz}(n,v,t) &=&\sum\limits_{l,j,s,p=1}^{4}\Re_{i}^{l}(n)\Re
_{j}^{\ast l}(n)\Re_{z}^{\ast s}(v)\Re_{p}^{s}(v)\Omega
_{iz}(n,v,0)e^{-it\{\lambda_{l}-\lambda_{s}\}},  \label{Coof}
\\
\qquad\quad\left|\Psi_{1}\right\rangle &=&|e_{1},e_{2},n-k\rangle, \quad
\left|\Psi_{2}\right\rangle =|g_{1}, e_{2},n\rangle,
\qquad \left|\Psi_{3}\right\rangle =|e_{1},g_{2},n\rangle ,
\nonumber
\\
\quad\qquad \left| \Psi_{4}\right\rangle &=&|g_{1},g_{2},n+k\rangle .  \nonumber
\end{eqnarray}
We denote by $\lambda_{m}$ the eigenvalues of the Hamiltonian $\widehat{H}
_{in}$ and $\Re_{i}^{l}(n)$ is the $i^{\underline{th}}$ element of the $l^{
\underline{th}}$ eigenvector. The coefficients $\Omega
_{iz}(n,v,0)$ specify the initial conditions for the field and atomic states. The eigenvalues $\lambda_{l}$ are to be found
from a fourth-order scalar equation, the roots of which may be easily
written in closed form for two identical atoms, arbitrary detuning in the
absence of the Stark shifts or for two nonidentical qubits at exact
resonance. For a general multiphoton interaction in the presence of
Stark shifts and detuning parameter, this problem can be treated
numerically. If we consider the dispersive approximation in which $\Delta
>>\mu_{n}\gamma ,$ (where $\gamma =\gamma_{2}\Delta $ is the dipole
coupling constant) and the second qubit is weakly coupled to the field, an
explicit expression for the final state $\mathcal{E}_{t}^{\ast }\rho $ can
be easily obtained {[37,38]}. In that case, it is straightforward to obtain
explicit expressions for $\Omega_{iz}(n,v,t)$, namely,
\begin{eqnarray}
\Omega_{11} &=&A_{n}^{t}A_{l}^{t*}b_{n-k}b_{l-k}\cos^{2}\theta_{1}\cos
^{2}\theta_{2}+B_{n}^{t}B_{l}^{t*}b_{n}b_{l}\sin^{2}\theta_{1}\cos
^{2}\theta_{2},
 \nonumber
\\
\Omega_{12} &=&A_{n}^{t}B_{l}^{t*}b_{n-k}b_{l-k}\cos^{2}\theta_{1}\cos
^{2}\theta_{2}-B_{n}^{t}A_{l}^{t}b_{n}b_{l}\sin^{2}\theta_{1}\cos^{2}\theta_{2},
\nonumber
\\
\Omega_{22} &=&A_{n}^{t*}A_{l}^{t}b_{n}b_{l}\sin^{2}\theta_{1}\cos
^{2}\theta_{2}+B_{n}^{t}B_{l}^{t*}b_{n-k}b_{l-k}\cos^{2}\theta_{1}\cos
^{2}\theta_{2},
 \nonumber
 \\
\Omega_{33} &=&
A_{n+k}^{t}A_{l+k}^{t*}
b_{n}b_{l}\cos^{2}\theta_{1}\sin
^{2}\theta_{2}+
B_{n+k}^{t}B_{l+k}^{t*}
 b_{n+k}b_{l+k}\sin^{2}\theta_{1}\sin^{2}\theta_{2},
\nonumber
\\
\Omega_{34} &=&-
A_{n+k}^{t*}B_{l+k}^{t}
b_{n}b_{l}\cos^{2}\theta
_{1}\sin^{2}\theta_{2}-
B_{n+k}^{t}A_{l+k}^{t*}
b_{n+k}b_{l+k}\sin^{2}\theta_{1}\sin^{2}\theta_{2},
\nonumber
\\
\Omega_{44} &=&
A_{n+k}^{t}A_{l+k}^{t*}
b_{n-k}b_{l-k}\sin^{2}\theta
_{1}\sin^{2}\theta_{2}+
B_{n+k}^{t}B_{l+k}^{t*}
 b_{n}b_{l}\cos^{2}\theta_{1}\sin^{2}\theta_{2},  \label{Coof-d}
\end{eqnarray}
where
\begin{eqnarray}
A_n^t &=&\frac{1}{2}\exp \left( -it\{g_{n}-\mu_{n}\}\right) \left\{ 1+
\frac{\gamma_{2}}{2\mu_{n}}+\exp \left( -2i\mu_{n}t\right) \left( 1-\frac{
\gamma_{2}}{2\mu_{n}}\right) \right\} ,
\nonumber
\\
B_n^t &=&-\frac{\gamma_{1}}{2\mu_{n}}\sqrt{\frac{n!}{(n-k)!}}\exp \left(
-it\{g_{n}-\mu_{n}\}\right) \left\{ 1-\exp \left( -2i\mu_{n}t\right)
\right\} ,
\\
\mu_{n} &=&\sqrt{\left( \gamma_{2}^{2}/4\right) +\frac{\gamma
_{1}^{2}(n+k)!}{n!}},\qquad g_{n}=\Delta +\gamma_{2}\left( n+\frac{k}{2}
\right),
\nonumber
\end{eqnarray}
$\Omega_{ij}=\Omega_{ji}^{\ast },$ $\Omega_{13}=\Omega_{14}=\Omega
_{23}=\Omega_{24}=0.$

A downside of analyzing more complex atomic system is that analytic
expressions for the final state functions and, consequently, for the matrix
elements are not always available. Therefore, the greatest benefit of
equation (10) which represents an analytical solution of the final state of
the system for this general model, is that it directly yields a method for
actually calculating any property related to the system.


\section{Atomic inversion and field properties}


We mainly devote the present section to consider the atomic inversion from
which the phenomenon of collapse and revival can be observed [18], and see
how it is affected in the present model. The population inversion
expressions for each qubit can be written as
\begin{equation}
\langle \sigma^{(i)}(t)\rangle =\frac{1}{2}Tr\left\{ \left|
e_{i}\right\rangle \left\langle e_{i}\right| -\left| g_{i}\right\rangle
\left\langle g_{i}\right| \right\} \rho_{i}(t),\qquad i=1,2,
\end{equation}
where $\rho_{i}(t)$ is the reduced atomic density matrix of the $i^{
\underline{th}}$ qubit which can be obtained by tracing out the field
variables i.e.,
\begin{equation}
\rho_{i}(t)=Tr_{j}\left( \mathcal{E}_{t}^{\ast }\rho \right) .
\end{equation}
In this case, the total atomic population inversion is given by
\begin{equation}
\langle \sigma (t)\rangle =\frac{1}{2}\left( \langle \sigma^{(1)}(t)\rangle
+\langle \sigma^{(2)}(t)\rangle \right) .  \label{totalAI}
\end{equation}
\begin{figure}[tbph]
\begin{center}
\includegraphics[width=7cm]{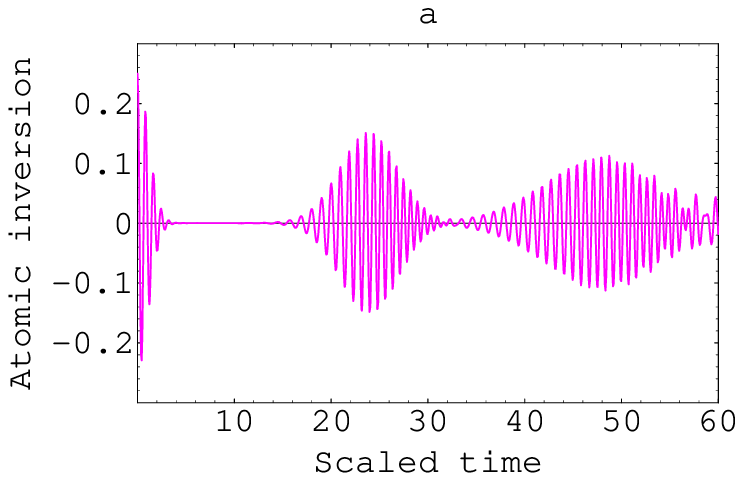}
\includegraphics[width=7cm]{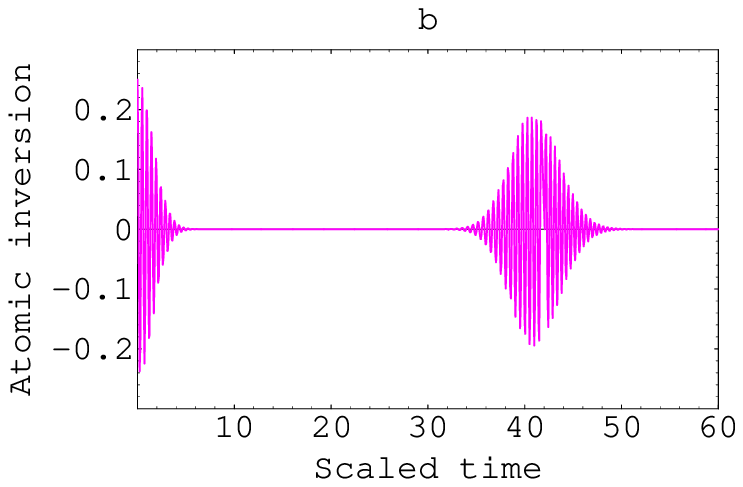}
\end{center}
\caption{Time history of the total atomic inversion $\langle\sigma(t)\rangle$ as a function of the scaled time $\gamma
_{1}t $. Calculations assume that $k=1,\theta_{2}=\pi /4,
\theta_{1}=0,$ $m=70,\gamma_{2}/\gamma_{1}=0.2$,
the detuning parameter $\Delta $ has zero value, and for different values of
$\eta $ where (a) $\eta =0.2$ and (b) $\eta =0.7$. }
\end{figure}
Figure 2 is a time history of the total atomic inversion $\langle \sigma
(t)\rangle $ according to equation (\ref{totalAI}) against the scaled time $
\gamma_{_{1}}t.$ In this figure we assume that the one-photon transition $
k=1,$ the second atom starts from a maximum entangled state ($\theta_{2}=
\frac{\pi }{4})$ and the first atom starts from a pure state $\theta_{1}=0,$
while initial field is a binomial state with $m=70$, and for different
values of the parameter $\eta ,$ where $\eta =0.2$ for figure 2a and $\eta
=0.7$ for figure 2b. The detuning parameter $\Delta $ has a zero value and $
\frac{\gamma_{2}}{\gamma_{1}}=0.2$. In this case and with a small value of
$\eta $ (say $\eta =0.2)$ the total atomic inversion exhibits well-known
collapses and revivals. However, any change of the binomial parameter $\eta $
leads to changing in the atomic inversion and consequence, increasing the
parameter $\eta ,$ leads to elongating the revival time while the atomic
inversion oscillates around the same value (see figure 2b). In general when $
\eta $ increases, the number of isolated revivals decreases at the same
period of time.
\begin{figure}[tbph]
\begin{center}
\includegraphics[width=7cm]{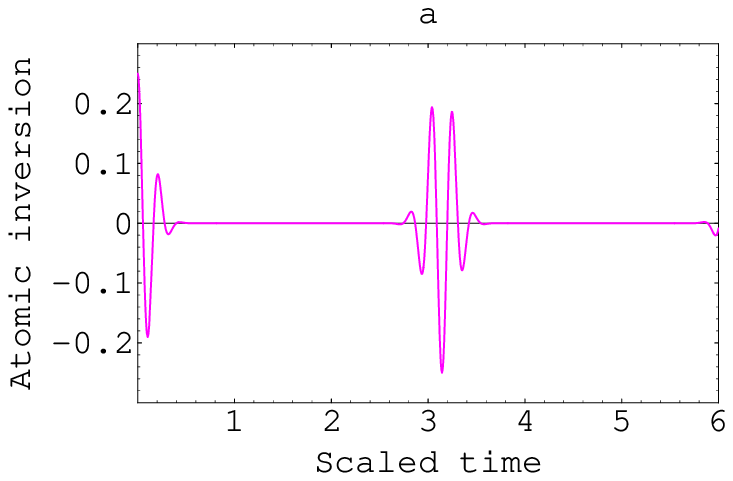}
\includegraphics[width=7cm]{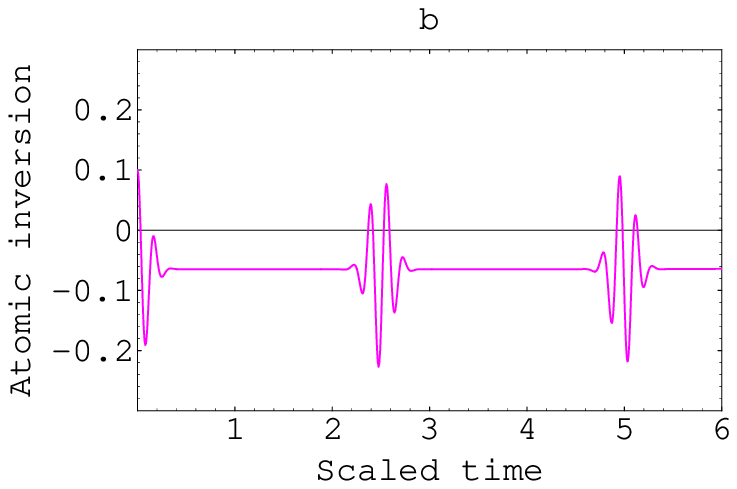}
\end{center}
\caption{ The evolution of the atomic inversion $\langle\sigma(t)\rangle$
as a function of the scaled time $\gamma
_{1}t $. Calculations assume that $k=2$ (the two-photon processes), $
\theta_{2}=\pi /4,\theta_{1}=0,$ $m=70,\gamma_{2}/
\gamma_{1}=0.2$, the detuning parameter $\Delta $ has zero value,
and for different values of $\eta $ where (a) $r=1,\eta =0.2$
and (b) $r=0.7,\eta =0.2$. }
\end{figure}
\begin{figure}[tbph]
\begin{center}
\includegraphics[width=7cm]{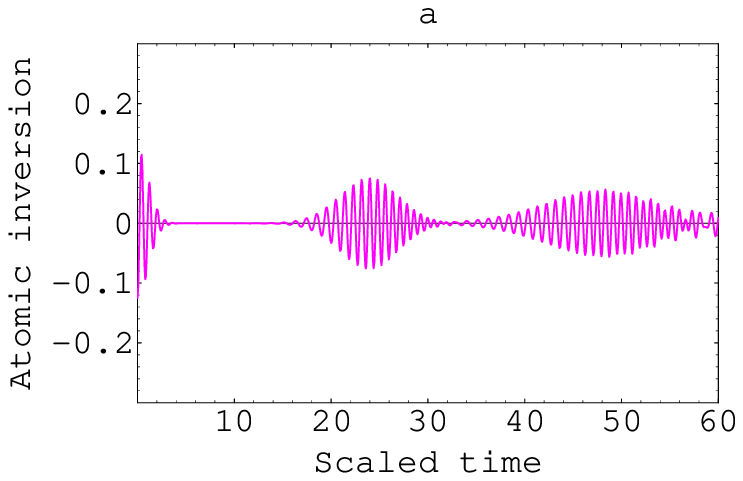}
\includegraphics[width=7cm]{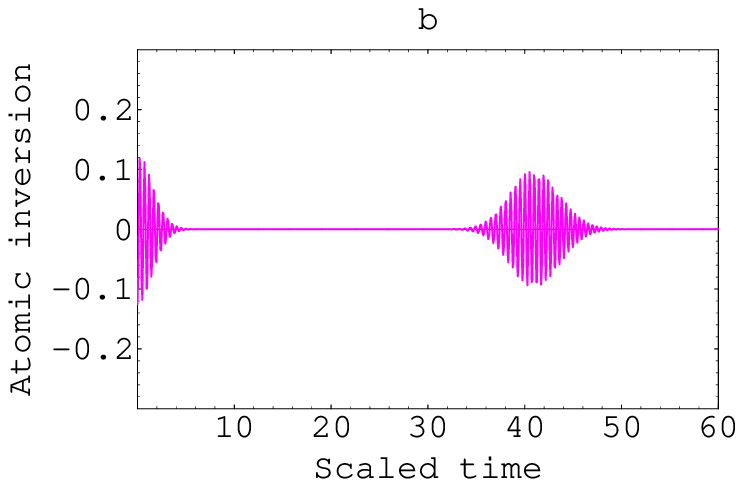}
\end{center}
\caption{The evolution of the atomic inversion $\langle \sigma (t)\rangle$
as a function of the scaled time $\gamma
_{1}t $. Calculations assume that $k=1,\theta_{2}=\pi /4,
\theta_{1}=\frac{\pi }{3},$ $m=70,\gamma_{2}/
\gamma_{1}=0.2$, the detuning parameter $\Delta $ has zero value,
and for different values of $\eta $ where (a) $\eta =0.2$
and (b) $\eta =0.7$. }
\end{figure}
A natural next question would be: Given a quantum state of which one knows
that it is entangled, how can $\langle \sigma (t)\rangle $ be affected by a
multiphoton transition$?$ \textrm{One may envision, for example, the
situation that a two-photon process is involved (i.e. $k=2)$.} In figure 3,
we consider this case in which the Stark shift will be taken into account
(we set $k=2$ and the other parameters are the same as in figure 2). We see
that the total atomic inversion has rapid oscillations with a periodical
collapses and revivals. This discussion has clearly demonstrated that the
general behavior of the total atomic inversion, when the two-photon
transition is involved, is remarkably affected by changing the number of
quanta and $\langle \sigma (t)\rangle $ is almost periodic. This periodicity
follows a consequence of the fact that the generalized Rabi frequency in the
two-photon transition is proportional to $n$ rather than to $\sqrt{n}$ which
is the case in the single-photon process. In the presence of the Stark
shifts namely $r=\sqrt{\beta_{2}^{(i)}/\beta_{1}^{(i)}},$ we note that the
Rabi frequency as well as the temporal width of the oscillations packets
decrease (see figure 3b). The most obvious difference, is that the
population inversion collapses to a nonzero value with the inclusion of
Stark shifts but to a zero value without it. This fact highlights an
important difference between the two situations.
\begin{figure}[tbph]
\begin{center}
\includegraphics[width=7cm]{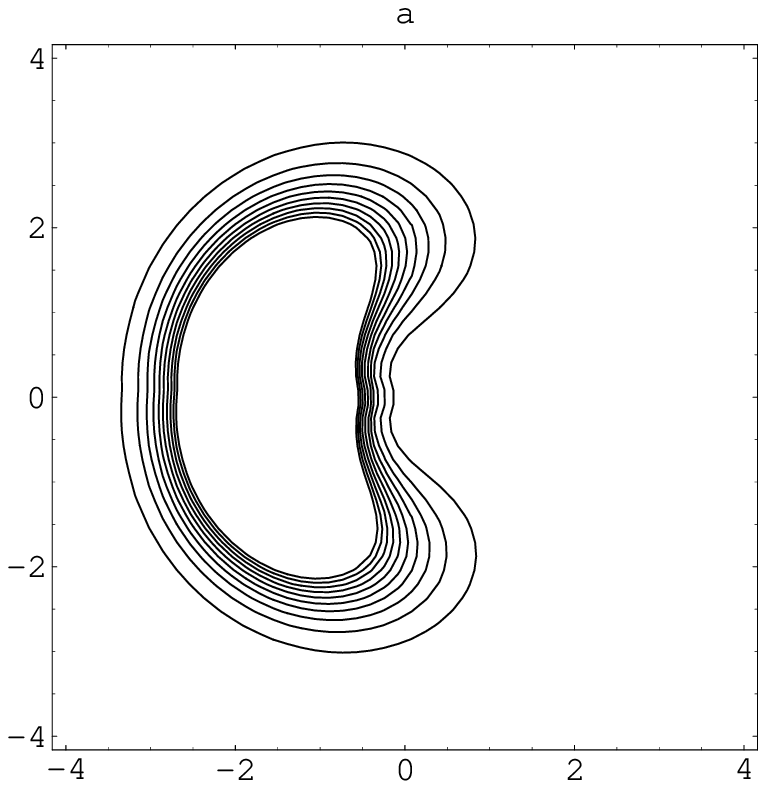}
\includegraphics[width=7cm]{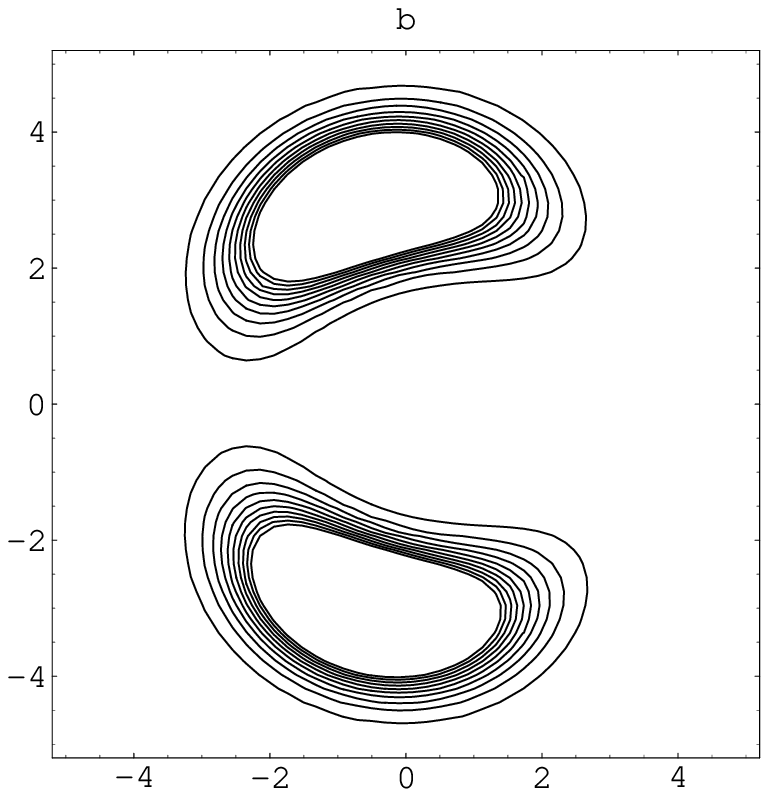}
\end{center}
\caption{Contour plot of the field reduced density matrix quasiprobability
distribution function against $X=Re.(\beta )$ and $Y=Im.(
\beta )$. Calculations assume that $k=1,\theta_{2}=\pi /4,
\theta_{1}=0,$ $m=70,\gamma_{2}/\gamma_{1}=0.2$,
and the detuning parameter $\Delta $ has zero value, where (a) $
\gamma_1t=\frac{5\pi }{2},\eta =0.2$ and (b) $\gamma
_1t=\frac{5\pi }{2},\eta =0.7$. }
\end{figure}

In what follows we shall choose the parameters to show the first atom starts
also from a mixed state, i.e $\theta_{1}=\frac{\pi }{3}$. In this case the
evolution of the atomic inversion still exhibits Rabi oscillations, showing
collapses and revivals. The oscillations are independent of $\theta_{1},$
but $\theta_{1}$ strikingly affects the amplitude of the inversion (see
figure 4). As soon as we take $\theta_{1}$ effect into consideration it is
easy to realize that the amplitude of the oscillations is decreased.
Furthermore if we take larger values of the binomial parameter $\eta $, the
revival time is elongated and the atomic inversion oscillates around zero.

At this end and after discussing a particular aspect of the atomic inversion
in the multi-photon two two-level qubits model with the initial field
prepared in the binomial state, which clearly exhibits the
collapses-revivals phenomenon and provide us with information about the
discrete nature of the quantum qubit-cavity field eigenvalues, we are going
to focus our attention on the representation of the field in phase space
which provides some aspects of the field dynamics. Perhaps the most
convenient quasiprobability to be used in this kind of problem is the $Q-$
function. The first step to be taken is the calculation of the reduced
density operator for the field $\rho_{F}(t)$, and then we get the $Q-$
function as
\begin{equation}
Q(x,y,t)=\frac{1}{\pi }\langle \zeta |\rho_{F}(t)|\zeta \rangle ,
\label{28}
\end{equation}
where $|\zeta \rangle $ is a coherent state with amplitude $\zeta =x+iy.$
The $Q-$function is not only a convenient tool to calculate expectation
values of anti-normally ordered products of operators, but also gives us a
new insight into the mechanism of interaction in the model under
consideration. Figure 5 shows a contour plot of the reduced density matrix $
Q-$function quasiprobability for different values of the binomial parameter $
\eta .$ When $\eta $ increases, the $Q-$function bifurcates into two blobs
rotating in the complex coherent state parameter plane in the clockwise and
counterclockwise directions with the same speed. The collision of the blobs
corresponds to the revival. This can be clearly seen if we compare the
numerical calculations for inversion and $Q-$function.


\section{Concurrence}


The characterization and classification of entanglement in quantum mechanics
is one of the cornerstones of the emerging field of quantum information
theory. Although an entangled two-qubit state $\mathcal{E}_{t}^{*}\rho $ is
not equal to the product $\mathcal{E}_{t}^{*}\rho_{1}$ and $\mathcal{E}
_{t}^{*}\rho_{2}$ of the two single-qubit states contained in it, it may
very well be a convex sum of such products. In general it is known that
microscopic entangled states are found that to be very stable, for example
electron-sharing in atomic bonding and two-qubit entangled photon states
generated by parametric down conversion. Entanglement as one of the most
nonclassical features of quantum mechanics is usually arisen from quantum
correlations between separated subsystems which can not be created by local
actions on each subsystem.

As more and more experimental realizations of entanglement sources become
available, it is necessary to develop different methods of measuring the
entanglement produced by different sources {[39]}. An ensemble is specified
by a set of pairs $\{(p_{i},|\psi_{i}\rangle )\}_{i=1}^{N}$, consisting of $
N$ state vectors $|\psi_{i}\rangle $ and associated statistical weights $
p_{i}$, and, $N$ is called the cardinality of the ensemble. The concurrence
of a bipartite state (i.e., a state over the bi-partite Hilbert space $\frak{
S}_{A}\otimes \frak{S}_{F}$) is defined by an almost magic formula {[30,31]}
\begin{equation}
C_{\mathcal{E}_{t}^{\ast }\rho }\left( t\right) =\max \left\{ 0,\lambda
_{1}-\lambda_{2}-\lambda_{3}-\lambda_{4}\right\} .
\end{equation}
We denote by $\lambda_{i}$ the square roots of the eigenvalues of $\left(
\mathcal{E}_{t}^{\ast }\rho \right) \times \widetilde{\left( \mathcal{E}
_{t}^{\ast }\rho \right) }$ in descending order, where
\begin{equation}
\widetilde{\left( \mathcal{E}_{t}^{\ast }\rho \right) }=\left( \sigma
_{y_{1}}\otimes \sigma_{y_{2}}\right) \left( \mathcal{E}_{t}^{\ast }\rho
\right)^{\ast }\left( \sigma_{y_{1}}\otimes \sigma_{y_{2}}\right) .
\end{equation}
where $\sigma_{y_{i}}$ is the Pauli matrix. The importance of this measure
follows from the direct connection between concurrence and entanglement of
formation$.$ It has been shown that {[30]} the entanglement of formation of
an arbitrary state $\mathcal{E}_{t}^{\ast }\rho $ is related to the
concurrence $C_{\mathcal{E}_{t}^{\ast }\rho }\left( t\right) $ by a function
\begin{equation}
E_{F}\left( C_{\mathcal{E}_{t}^{\ast }\rho }\left( t\right) \right) =\pi
^{+}(t)\log \pi^{+}(t)+\pi^{-}(t)\log \pi^{-}(t),
\end{equation}
where
\begin{equation}
\pi^{\pm }(t)=\frac{1}{2}\left( 1+\sqrt{1-C_{\mathcal{E}_{t}^{\ast }\rho
}^{2}\left( t\right) }\right) .
\end{equation}
The entanglement of formation is monotonically increasing with respect to
the increasing concurrence. If we could write the general bipartite pure state as [40,31]
\begin{eqnarray}
\left| \psi (t)\right\rangle \equiv
\sum\limits_{i,j=0}^{d-1}\sum\limits_{n=0}^{\infty }\wp_{ij}(n,t)\left|
i,j,n\right\rangle,
\end{eqnarray}
in this case, the concurrence can be calculated as
\begin{eqnarray}
C_{\mathcal{E}_{t}^{\ast }\rho }\left( t\right) &=&\left\{ 2\left[
1-Tr\left( \rho_{A}^{2}(t)\right) \right] \right\}^{\frac{1}{2}}  \nonumber
\\
&=&\left\{ \sum\limits_{n=0}^{\infty }\sum\limits_{i,j,l,k=0}^{d}\left| \wp
_{ik}(n,t)\wp_{jm}(n,t)-\wp_{im}(n,t)\wp_{jk}(n,t)\right|^{2}\right\}^{
\frac{1}{2}}.
\end{eqnarray}
The concurrence $C_{\mathcal{E}_{t}^{\ast }\rho }\left( t\right) $ as a
measure of the degree of entanglement ensures the scale between $0$ and $1$
and monotonously increases as entanglement grows. A note of caution about
how to interpret the state of a physical system in terms of quantum
entanglement may be in place here. The previous standard definitions of
quantum entanglement tacitly assume that, every state in the bipartite or
multipartite Hilbert space is in principle available as a physical state and
local as well as global quantum operations, measurements and unitary
transformations, can be performed on the Hilbert space.


\subsection{Results}


We now discuss applications of the above equation to specific situations. As
stated above, an important situation is that, when $C_{\mathcal{E}_{t}^{\ast
}\rho }\left( t\right) =0$ the two qubits are separable and $C_{\mathcal{E}
_{t}^{\ast }\rho }\left( t\right) =1$ indicates maximum entanglement between
the two qubits.
\begin{figure}[tbph]
\begin{center}
\includegraphics[width=8cm]{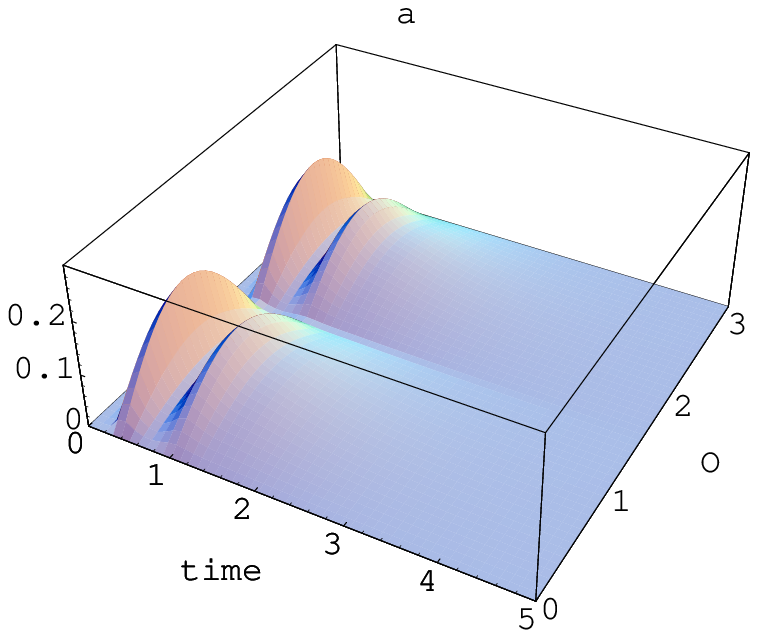} 
\includegraphics[width=8cm]{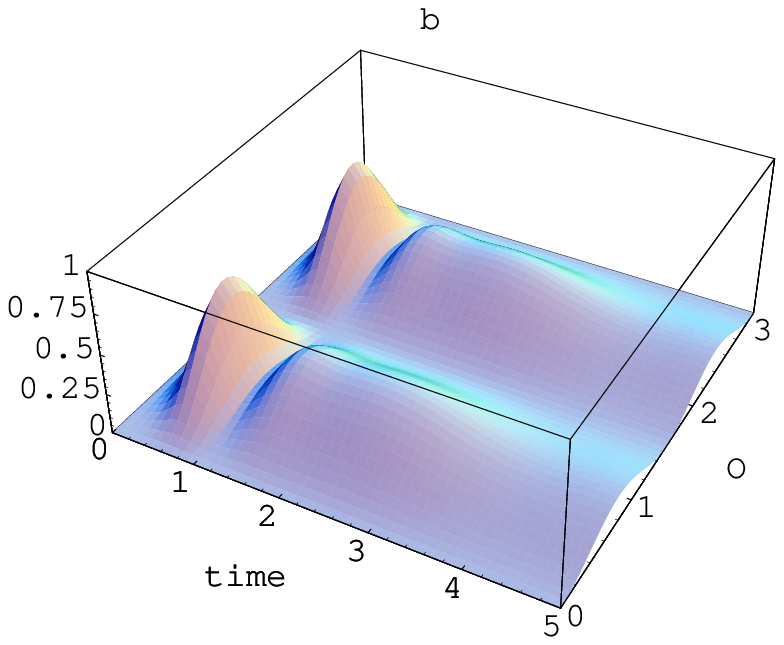}
\end{center}
\caption{Plot of the concurrence $C_{\mathcal{E}_{t}^{\ast }\rho
}\left( t\right) $ as a function of the scaled time $\gamma_{1}t$
and $\theta_2 $. The field initially in the binomial states with $
\eta =0$, $k=1,\Delta =0,\gamma_{2}/\gamma_{1}=0.2$
, and for different values of the mean photon number $\eta m=\bar{n}$
, where (a) $\bar{n}=20$ and (b) $\bar{n}=8$. }
\label{motion}
\end{figure}
In our numerical examples we have used physical parameters from some recent
experiments {[22, 41]}, but extrapolated the qubit transit time $t$ to
rather large values of $\gamma_{1}t$. It is, of course, an experimental
challenge to obtain a one-qubit source and atomic life-times of the atomic
states involved such that these large values of $\gamma_{1}t$ can be
reached. An interesting question is whether or not the
entanglement is affected by the different parameters of the present system
with the initial state in which one of the qubits is prepared in its excited
state and the other in the mixed state. In particular, the mixed state
parameter $\theta_2$, the scaled time $\gamma_{1}t,$ and the parameters from
the initial state of the field ($m,$ $\eta ).$ A numeric evaluation of the
concurrence as an entanglement measure leads to the plot in figures 6-10. We
now pause to touch on certain concurrence features when $\eta \approx 0$
(i.e., the coherent state)$,$ the mean photon number $\eta m=20$ for figure
6a and $\eta m=10$ for figure 6b. The maximum value of the entanglement
increased as the mean photon number is decreased (see figure 6), but the
entanglement vanish as the time goes on for large values of the mean photon
number this is not the case when $\overline{n}$ takes small values (see
figure 6b). It is interesting to note that, the maximum entanglement is
achieved when $\theta_2 =\frac{n\pi }{4},$ $n=\pm 1,\pm 3,\pm 5,..$ while $C_{
\mathcal{E}_{t}^{\ast }\rho }\left( t\right) \approx 0,$ for $\theta_2 =\frac{
m\pi }{2},$ $m=0,\pm 1,\pm 2,..$. Also, the entanglement shows symmetry
around $\theta_2 =0.$ Put differently, with different values of $\eta $, such
as $\eta $=0.7, one will have a very small amount of entanglement at the
initial period of time only and this amount disappear when the time goes on
(see figure 7a). Indeed, the comparison of plots figure 6a and figure 7a
where $\eta =0$ and $\eta =0.7$, respectively demonstrates that the
entanglement in both cases has somewhat similar behavior corresponding to
different values of $\theta_2.$ In accord with the initial conditions $\eta
=0.7$ and $m=70,$ figure 7b shows more oscillations at the same period of
time while the entanglement survive in this case longer. Also, small amount
of entanglement is repeated several times with the time development. At the
period $2\leq \gamma_{1}t\leq 3,$ the entanglement shows small oscillations
throughout the manipulations as a result of increasing the binomial
parameter $\eta =0.9$. The comparison between figure 7a and figure 7b shows
the obvious effects of the binomial parameter $\eta $ in the dynamics of the
system.
\begin{figure}[tbph]
\begin{center}
\includegraphics[width=8cm]{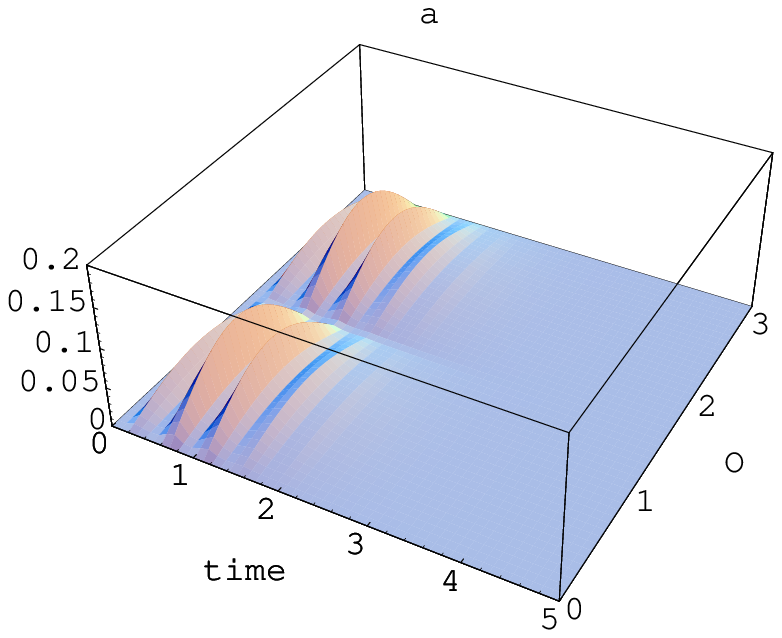} 
\includegraphics[width=8cm]{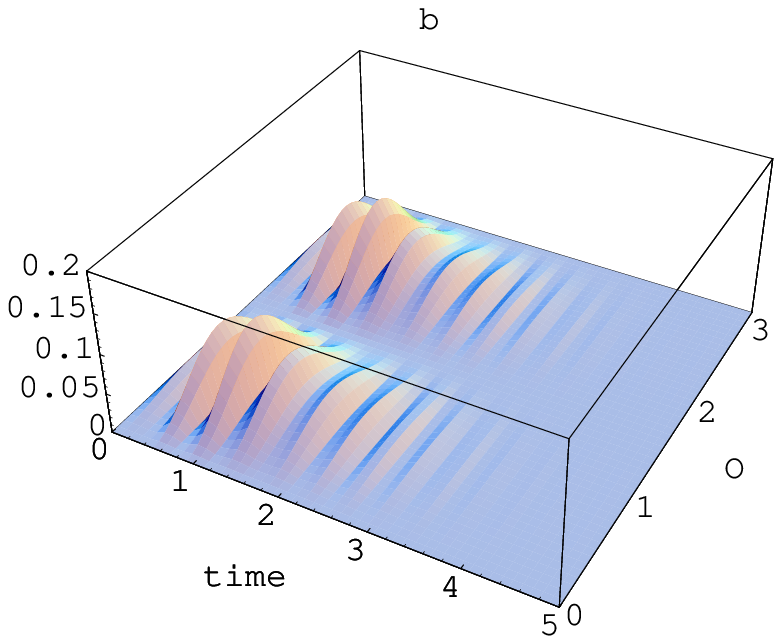}
\end{center}
\caption{ Plot of the concurrence $C_{\mathcal{E}_{t}^{\ast }\rho
}\left( t\right) $ as a function of the scaled time $\gamma_{1}t$
and $\theta_2 $. The field initially in the binomial states with $\bar{
n}=20$, $k=1,\Delta =0,m=70\gamma_{2}/\gamma_{1}=0.2$, and
for different values of $\eta $ where (a) $\eta =0.7$ and
(b) $\eta =0.9$. }
\label{motion}
\end{figure}
Next, we will analyze the influence of dispersive approximation on the
appearance or disappearance phenomenon of the entanglement previously found.
Dispersive effects can be conveniently incorporated by assuming that $\gamma
_{2}<<\gamma_{1},$ (such as $\frac{\gamma_{2}}{\gamma_{1}}=0.01)$. It is
remarkable to see that with $\eta =0.7$, entanglement is nearly washed out
for the initial stage of the interaction time. While more oscillations have
been observed when the time goes on (see figure 8a). The situation is
changed for $\eta =0.9$ (see figure 8b), in this case the maximum
entanglement increased further and start earlier. The initial period in
which the entanglement washed out in figure 8a decreased with increasing the
parameter $\eta $. Also, the maximum value of the entanglement is increased
with increasing $\eta .$
\begin{figure}[tbph]
\begin{center}
\includegraphics[width=8cm]{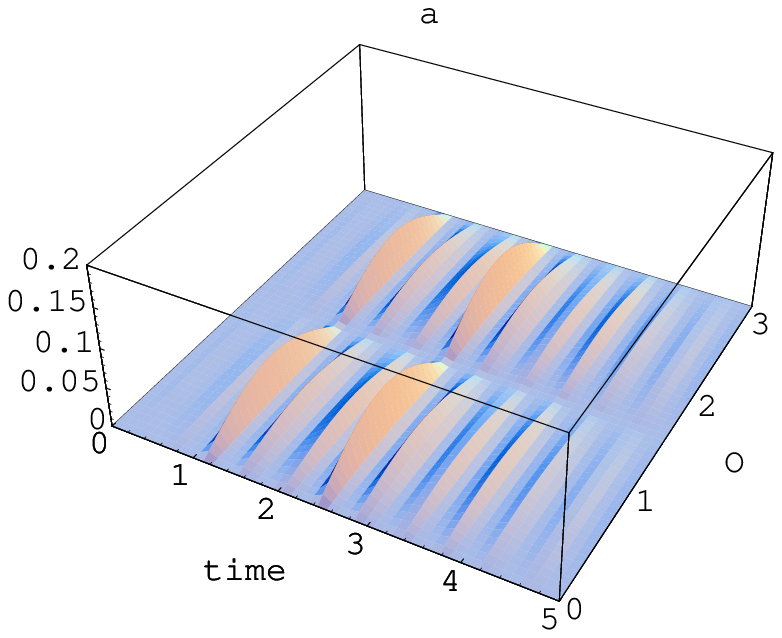} 
\includegraphics[width=8cm]{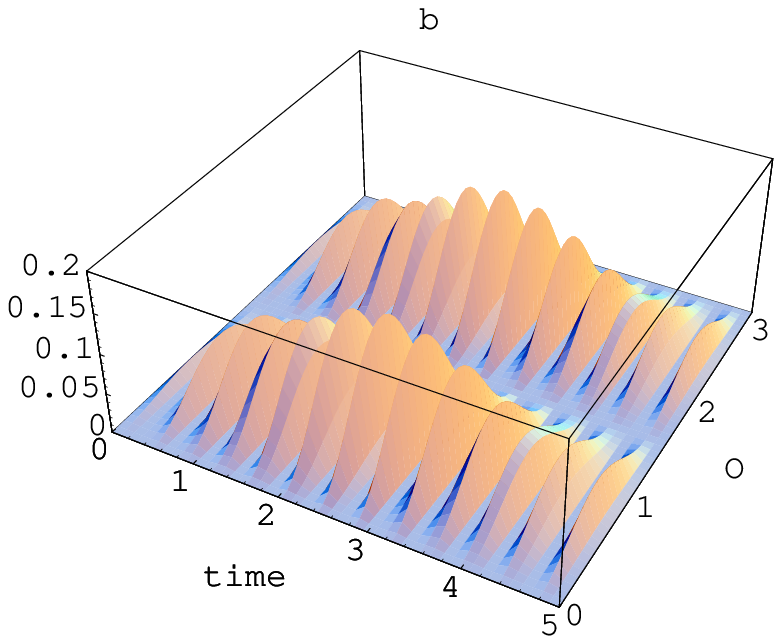}
\end{center}
\caption{ The same as figure 2 but $\gamma_{2}/\gamma
_{1}=0.01$. }
\label{motion}
\end{figure}
The number of quanta effect on the entanglement is particularly pronounced
as this number takes large values. In figure 9a the entanglement
is plotted as a function of $\gamma_{1}t$ and $\theta_2 $ for $k=2$. However,
we note from this plot that the maximum amount of the two-qubit entanglement
indeed does move closer to the point at $\gamma_{1}t=0$. Again we notice
some similarities with other plots in the sense that there is no
entanglement for $\theta_2=\frac{m\pi }{2}$ and maximum value at $\theta_2=
\frac{n\pi }{4},$ (see figure 9a)$.$ From our further calculations (which
were not presented here) it is clear that we can get the same amount of
entanglement using the present measure (concurrence) and the negativity as a
measure of entanglement, in agreement with our previous result {[42]}. Also,
the qubits and radiation subsystems exhibit alternating sets of collapses
and revivals due to the initial mixed states of the qubits and radiation
employed here. Let us now consider the situation when the qubits are
initially both in mixed state and nonzero detuning. It is surprising that a
nonresonant case (nonzero detuning) can entangle two qubits depending on
their atomic initial preparation. First of all, we note that the
entanglement appears only for the initial period of the interaction time$.$
These properties show that the role played by the detuning on the
entanglement is essential.

The above results pose two intriguing questions: (i) Why does the
entanglement become maximum for $\theta_2=\frac{n\pi }{4},$ and takes zero
value for $\theta_2 =0,$ $\frac{n\pi }{2}?$, (ii) Why does the entanglement
due to the concurrence behave essentially in a similar way with different
values of the mixed state parameter whatever values of the other parameters?
In what follows, we propose an analytic expressions which can give an answer
to these questions. We now first analyze the reason why the concurrence does
not exceed zero value when $\theta_2 =0.$ To prove that analytically, one may
first consider $\theta_{i}=0,$ then we have only nonzero values of $\Omega
_{11},\Omega_{12}$ and $\Omega_{22}$ and all the other $\Omega_{ij}$
vanish, i.e. the coefficients $\Omega_{ij}$ in equation (\ref{Coof-d})
reduce to
\begin{eqnarray}
\Omega_{11} &=&
A_{n}^{t}A_{l}^{t*}b_{n-k}b_{l-k},
\nonumber
\\
\Omega_{12} &=&A_{n}^{t}B_{l}^{t*} b_{n-k}b_{l-k},
\nonumber
\\
\Omega_{22} &=&B_{n}^{t}B_{l}^{t*}b_{n-k}b_{l-k}.
\end{eqnarray}
Using these coefficients with the above definition of the concurrence we
easily find that $C_{\mathcal{E}_{t}^{\ast }\rho }\left( t\right) =0,$ this
is also in agreement with the numerical calculations. On the other hand, if
we consider the first qubit in its excited state and the second qubit in a
mixed state, then equation (\ref{Coof-d}) reduces to$,$ \smallskip
\begin{eqnarray}
\Omega_{11} &=&A_{n}^{t}A_{l}^{t*}b_{n-k}b_{l-k}\cos^{2}\theta ,
\nonumber \\
\Omega_{12} &=&A_{n}^{t}A_{l}^{t*}b_{n-k}b_{l-k}\cos^{2}\theta ,
\nonumber \\
\Omega_{22} &=&B_{n}^{t}B_{l}^{t*}b_{n-k}b_{l-k}\cos^{2}\theta ,
\nonumber \\
\Omega_{33} &=&A_{n+k}^{t}A_{l+k}^{t*}b_{n}b_{l}\sin^{2}\theta ,
\nonumber \\
\Omega_{34} &=&-
A_{n+k}^{t}B_{l+k}^{t*}
b_{n}b_{l}\sin^{2}\theta ,
\nonumber \\
\Omega_{44} &=&B_{n+k}^{t}B_{l+k}^{t*}b_{n}b_{l}\sin^{2}\theta .
\end{eqnarray}
For simplicity we used $\theta $ instead of $\theta_{2}.$ In this case the
concurrence is given by
\begin{eqnarray}
C_{\mathcal{E}_{t}^{\ast }\rho }(t) &=&\left| \frac{\sin 2\theta }{2}\right|
\biggl\{\sum\limits_{n=0}^{\infty }b_{n}^{2}b_{n-k}^{2}\biggl|A_{n}^{t}B_{n+k}^{t*}
-B_{n}^{t}A_{n+k}^{t*}\biggr|^{2}\biggr\}^{\frac{1}{2}}.  \label{Conc-S}
\end{eqnarray}
\begin{figure}[tbph]
\begin{center}
\includegraphics[width=8cm]{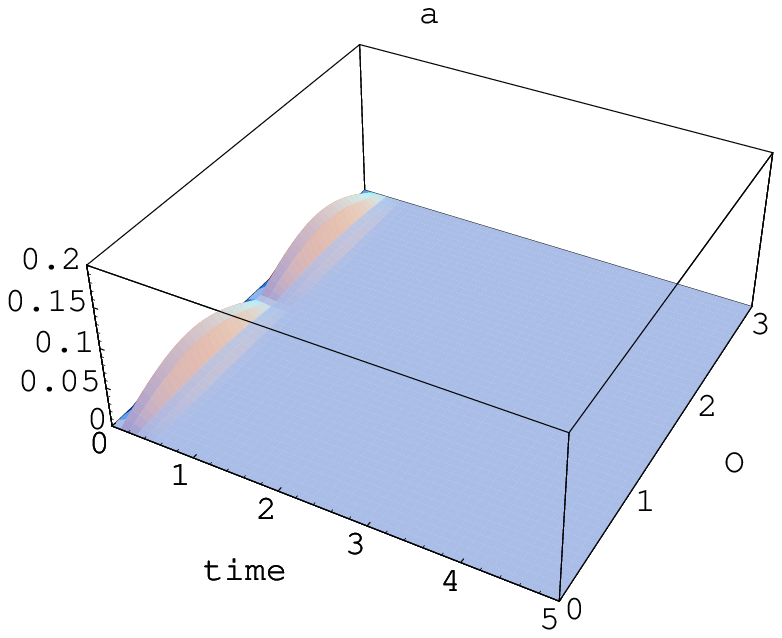}
\includegraphics[width=8cm]{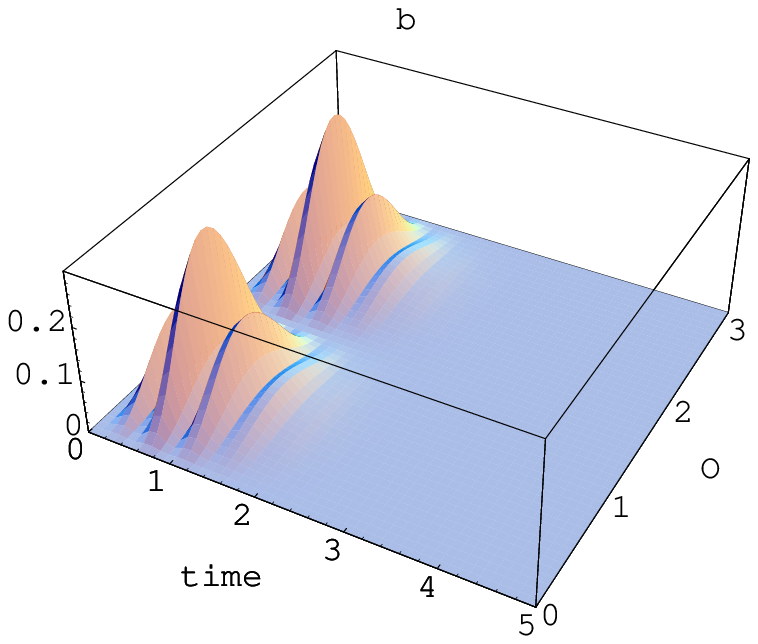}
\end{center}
\caption{Plot of the concurrence $C_{\mathcal{E}_{t}^{\ast }\rho
}\left( t\right) $ as a function of the scaled time $\gamma_{1}t$
and $\theta_2 $. The field initially in the binomial states with $\bar{
n}=20$, $m=70,\eta =0.7,\gamma_{2}/\gamma_{1}=0.2$
, where (a) $k=2,\Delta =0$ and (b) $k=1,\frac{\Delta }{\gamma_{1}}
=10$. }
\label{motion}
\end{figure}
From equation (\ref{Conc-S}) we see that for $\theta =\frac{n\pi }{4},$ we
will get maximum value of $C_{\mathcal{E}_{t}^{\ast }\rho }\left( t\right) $
and for $\theta =\frac{n\pi }{2},$ $C_{\mathcal{E}_{t}^{\ast }\rho }\left(
t\right) =0.$ It is also clearly from this equation, the concurrence depend
on the amplitude of the initial binomial state. Here, we clarify that it can
be done by using the different initial state, which is strongly affected by
the atomic number representation. This naturally leads to the use of
occupation numbers of different single-qubit basis states in quantifying
multi-qubit entanglement even when the number of qubits is conserved. The
occupation-numbers of different modes have already been used in quantum
computing {[43]. }It is important to refer here to the work in Ref. {[44]}
in which experimentally a superposition state of the ground state and a
non-maximally entangled antisymmetric state in two trapped ions has been
realized. In the experiment two trapped barium ions were sideband cooled to
their motional ground states. Transitions between the states of the ions
were induced by Raman pulses using co-propagating lasers. The non-maximally
entangled state was used {[45]} to demonstrate the intrinsic difference
between quantum and classical information transfers. The difference arises
from the different ways in which the probabilities occur and is particularly
clear in terms of entangled states.

\smallskip Before we conclude, it is necessary to give a brief discussion on
the experimental realization of the present model. It was reported that the
cavity can have a photon storage time of $T=1$ ms (corresponding to $
Q=3\times 10^{8}$ ). The radiative time of the Rydberg atoms with the
principle quantum numbers $49$, $50$ and $51$ is about $2\times 10^{-4}s$.
The coupling constant of the atoms to the cavity field is $2\pi \times 24$
kHz {[46].} \smallskip The experiment may be described schematically as
follows: a stream of pairs of rubidium atoms in circular Rydberg states,
time and velocity selected, was sent through a resonant cavity {[47]}. Each
pair consisted of one atom in a state with principle quantum number 51, and
the other in a state with principle quantum number 50, the frequency of
transition between the two states being 51.1 GHz. Also, there was an
experiment [47] in which the phase of oscillation of each of two entangled
qubits is individually controlled and can be adjusted to provide the
necessary values for testing a Bell inequality{.}


\section{Conclusion}


To sum up, in the main part of this paper we have investigated the
properties of the dynamically emerging entanglement in the
multiphoton two two-level qubits. We have treated the more general
case where initial states of the two qubits can be mixed with a
binomial state of the field. We have obtained an exact solution of
the density operator taking into account the dispersive limit, and
thus provides insight into the behavior of more complicated
two-qubit systems. We have investigated the interaction of the
binomial states (which reduce to number and coherent states in two
different limits) with atomic systems in the framework of the two
two-level qubits, and describe the response of the atomic system
as it varies between the Rabi oscillations and the
collapse-revival mode and investigate the total atomic inversion
and the quasiprobability distributions. The idea of using the
concurrence as an entanglement measure offers many attractive
features. Entanglement is measured via the concurrence, currently
used only for an arbitrary system of two qubits, but similar
analysis can in principle be applied to other systems such as a
bipartite system with arbitrary dimensions. As we anticipated,
this system exhibits some novel features in comparison with the
single qubit system. We have found that some different regimes
occur, depending on the actual initial joint product state, the
number of quanta and the binomial state parameters. It was
demonstrated that quantum entanglement is stored in the model
system considered, and that the nature of this entanglement is
strongly dependent on the detuning of the atomic levels, an effect
that may have important consequences in other nonlinear processes.


\begin{thebibliography}{99}
\bibitem{1}  Messikh A,  Ficek Z and  Wahiddin M R B 2003 J. Opt. B:
Quantum Semiclass. Opt. {\bf 5} L1;

 Zhou L,  Song H S and  Li C 2002 J. Opt. B: Quantum Semiclass. Opt. {\bf 4} 425;

 Bennett C H, Brassard G,  Crepeau C,  Jozsa R,  Peres A, and Wootters W K 1993
 Phys. Rev. Lett. {\bf 70} 1895.

\bibitem{2}   Bouwmeester D,  Pan J-W,  Mattle K,  Eibl M, Weinfurter H, and
Zeilinger A, Nature {\bf 390} 575;

 Kim M S, Jinhyoung Lee, Ahn D  and  Knight P L 2002 Phys. Rev. A {\bf 65} 040101.

\bibitem{3} Deutsch D and  Jozsa R 1992 Proc. R. Soc. Lond. A {\bf 439} 553.

\bibitem{4} Abdel-Aty M 2000 J. Phys. B: At. Mol. Opt. Phys. {\bf 33} 2665;

Abdel-Aty M, and Obada A-S F 2003 Eur. Phys. J. D {\bf 23} 155;

Simon D R  1997 SIAM J. Comput. {\bf 26} 1474.

\bibitem{5}  Grover L K 1997 Phys. Rev. Lett. {\bf 79} 325;

Senitzky I R 2002 J. Phys. B: At. Mol. Opt. Phys. {\bf 35} 3029.

\bibitem{6}   Rudolph O 2001 J. Math. Phys. {\bf 42} 5306;

 Williams C P and  Clearwater S H 1998 Explorations in Quantum
 Computing (New York: Telos, Springer-Verlag).

\bibitem{7}  Abdel-Aty M 2003 J. Math. Phys. {\bf 44} 1457 and Virtual J.
Quant. Infor. April 2003, Volume 3, Issue 4.

\bibitem{8}   Sorensen A and Molmer K 2000 Phys. Rev. A {\bf 62} 022311.

\bibitem{9}  Horodecki M, Phy. Rev. A, {\bf 57} 3364 (1998).

\bibitem{10} Bennett C H 1995 Phys. Today {\bf 48} 24.

\bibitem{11} Terhal M, Horodecki M,  Leung D W,  DiVincenzo D P
2002 J. Math. Phys. {\bf 43} 4286.

\bibitem{12} Rudolph O 2001 J. Math. Phys. {\bf 42} 5306.

\bibitem{13} Barenco A,  Bennett C H,  Cleve R,  DiVincenzo D P,
Margolus N ,  Shor P,  Sleator T,  Smolin J A and  Weinfurter H
1995 Phys. Rev. A {\bf 52} 3457.

\bibitem{14} Iwai T  and  Hirose T 2002 J. Math. Phys. {\bf 43} 2907.

\bibitem{15} Ficek Z and  Tanas R 2002 Phys. Rep. {\bf 372} 369.

\bibitem{16} Agarwal G S 1974, Quantum Optics, Springer Tracts in Modern
Physics Vol. 70 (Springer-Verlag, Berlin).

\bibitem{17} Milman P, and Mosseri R arXiv: quant-ph/0302202;

 Verstraete F and Verschelde H 2002 Phys. Rev. A {\bf 66} 022307.

\bibitem{18} Kudryavtsev I K,  Lambrecht A,  Moya-Cessa H and Knight
P L 1993 J. Mod. Opt. {\bf 40} 1605;

Obada A-S F  and  Omar Z M 1993 J. Egypt. Math. Soc. {\bf 1} 63.

\bibitem{19}  Jex I 1990 Quantum Optics {\bf 2} 433; 1990 J. Mod. Opt.
{\bf 39} 835.

\bibitem{20}   Song T Q,  Feng J,  Wang W Z and  Xu J Z 1995 Phys. Rev. A
{\bf 51} 2648.

\bibitem{21}  Kudryavtsev I K  and Knight P L 1993 J. Mod. Opt. {\bf 40}
1673.

\bibitem{22} Tittonen I,  Stenhplm S and Jex I 1996 Opt. Commun. {\bf 124} 271.

\bibitem{23} Abdel-Aty M 2003 J. Opt. B: Quantum Semiclass. Opt. {\bf 5}
349.

\bibitem{24} Ashraf M M 1999 Opt. Commun. {\bf 166} 49.

\bibitem{25}  Ashraf  I and  Toor A H 2000 J. Opt. B: Quantum Semiclass. Opt.
{\bf 2} 772;

\bibitem{26}  Ashraf M M 2001 J. Opt. B: Quantum Semiclass. Opt. {\bf 3} 39.

\bibitem{27}  Yu Shi 2003 Phys. Rev. A {\bf 67} 024301.

\bibitem{28} Tessier T E,  Deutsch I H, Delgado A  and
Fuentes-Guridi I 2003 arXiv:quant-ph/0306015.

\bibitem{29} Kim M S,  Lee J,  Ahn D and Knight P L 2002 Phys. Rev.
A {\bf 65} 040101.

\bibitem{30} Wootters W K 1998 Phys. Rev. Lett. {\bf 80,} 2245.

\bibitem{31} Hill S and  Wootters W K 1997 Phys. Rev. Lett. {\bf 78} 5022;

 Hughston L,  Jozsa R and Wootters W 1993 Phys. Lett. A {\bf 183} 14;

 Lozinski A,  Buchleitner A, Zyczkowski K  and  Wellens T 2003 Europhys. Lett.
{\bf 62} 168;

 Rungta P,  Bu\v{z}ek V,  Caves C M,  Hillery M and  Milburn G J. 2001 Phys.
Rev. A {\bf 64} 042315.

\bibitem{32} Fan H,  Matsumoto K and Imai H 2003 J. Phys. A: Math. Gen.
{\bf 36} 4151.

\bibitem{33} Stoler D and  Saleh B E A and  Teich M C 1985 Opt. Acta
{\bf 32} 345.

\bibitem{34} Mahran M H,  Abdalla M S, Obada A-S F  and
El-Orany F A A
 1998 Nonlinear Opt. {\bf 19} 189.

\bibitem{35} Vidal G and  Tarrach R 1999 Phys. Rev. A {\bf 59} 141.

\bibitem{36} Shor P W 2002 J. Math. Phys. {\bf 43} 4334.

\bibitem{37} Brune M,  Haroche S,  Raimond J M,  Davidovich L and
Zagury N 1992 Phys. Rev. A {\bf 45} 5193.

\bibitem{38} Roversi J A, Vidiella-Barranco A  and  Moya-Cessa H 2003 Mod.
Phys. Lett. B {\bf 17} 219.

\bibitem{39} Mair A,  Vaziri A,  Weihs G, and  Zeilinger A 2001 Nature {\bf
412} 313.

\bibitem{40} Albererio S and  Fei S M 2001 J. Opt. B: Quant. Semiclass. Opt.
{\bf 3} 223.

\bibitem{41} Benson O,  Raithel G and Walther H 1994 Phys. Rev. Lett. {\bf 72}
3506 and Dynamics of the Micromaser Field in Electron Theory and
Quantum Electrodynamics: 100 Years Later, Ed.  Dowling J P 1997
(Plenum Press, New York).

\bibitem{42} Obada A-S F  and Abdel-Aty M 2003 J. Math. Phys. preprint.

\bibitem{43} E Knill, R Laflamme and  Milburn G J 2001 Nature (London)
{\bf 409} 46.

\bibitem{44} Turchette Q A,  Wood C S,  King B E,  Myatt C J,
Leibfried D ,  Itano W M,  Monroe C and  Wineland D J 1998 Phys.
Rev. Lett. {\bf 81} 3631.

\bibitem{45} Franke S,  Huyet G, and  Barnett S M 2000 J. Mod. Opt. {\bf 47} 145.

\bibitem{46} Brune M, E Hagley,  Dreyer J,  Maitre X,  Maali A,
Wunderlich C ,  Raimond J M, and  Haroche S 1996 Phys. Rev. Lett.
{\bf 77} 4887.

\bibitem{47} Hagley E,  Maitre X,  Nogues G,  Wunderlich C,  Brune M,  Raimond J
M and  Haroche S 1997 Phys. Rev. Lett. {\bf 79} 1.

\bibitem{48} Rowe M A,  Kielpinski D,  Meyer V,  Sackett C A,
Itano W M,  Monroe C, and  Wineland D J 2001 Nature {\bf 409} 791.
\end{thebibliography}
\end{document}